SOFTWARE

# MIMTool: A Tool for Drawing Molecular Interaction Maps

Mine Edes[1], Can Özturan[*2], Türkan Haliloğlu[3], Augustin Luna[4] and Ruth Nussinov[5]

**Abstract**

**Background**: To understand protein function, it is important to study protein-protein interaction networks. These networks can be represented in network diagrams called protein interaction maps that can lead to better understanding by visualization. We address the problem of drawing of protein interactions in Kohn's Molecular Interaction Map (MIM) notation. Even though there are some existing tools for graphical visualization of protein interactions in general, there is no tool that can draw protein interactions with MIM notation with full support.

**Results:** MIMTool was developed for drawing protein interaction maps in Kohn's MIM notation. MIMTool was developed using the Qt toolkit libraries and introduces several unique features such as full interactivity, object dragging, ability to export files in MIMML, SBML and line drawing with automatic bending and crossover minimization, which are not available in other diagram editors. MIMTool also has a unique orthogonal edge drawing method that is both easy and more flexible than other orthogonal drawing methods present in other interaction drawing tools.

**Conclusions:** MIMTool facilitates faster drawing, updating and exchanging of MIMs. Among its several features, it also includes a semi-automatic drawing algorithm that makes use of shortest path algorithm for constructing lines with small number of bends and crossings. MIMTool contributes a much needed software tool that was missing and will facilitate wider adoption of Kohn's MIM notation.

[*]Correspondence: ozturaca@boun.edu.tr

[1] Computer Engineering Department, Bogazici University, Bebek 34342 Istanbul, Turkey

Full list of author information is available at the end of the article

## Introduction

Cancer, one of the most common and lethal diseases of today, is mainly caused by malfunctioning of tumor suppressor proteins. The most widely studied tumor suppressor, which is mutated in half of reported cancer cases, is p53; another, which has common functions with and is a homologue of p53, is p73 [1]. To understand the functioning of these proteins, it is important to understand the protein-protein interaction networks they participate in. To facilitate this understanding, interactions of proteins can be represented as pathway diagrams. These diagrams can facilitate easy analysis of the protein interactions, which in turn can help guide the formation of experiments and help researchers understand the wider context of a system thereby having impact on cancer research, drug design and protein engineering.

There are several proposals for specific notations for the visualization of the protein interactions meant to drive their standardization. They include Pirson [2], Cook [3], Kurata [4], Kitano [5], and Kohn [6]. There are some tools for graphical visualization of the protein interaction networks. However, we are not aware of any tool that was specifically designed to fully support Kohn's Molecular Interaction Map (MIM) notation. This paper introduces MIMTool; a tool for semi-automatic drawing of protein interaction maps in Kohn notation. MIMTool has been developed as part of a multi-year project as an open-source package for use by the researchers and is available at http://code.google.com/p/mimtool/. MIMTool was developed in C++ using the Qt graphical user interface (GUI) framework. MIMTool's features include an XML-based format for MIM storage, called MIMML [7], the ability to export Systems Biology Markup Language (SBML) [8] and the ability to export maps in PNG and PDF formats.

## Background

There are some tools available for visually modeling 2D and 3D protein-protein interactions. These are: Interviewer [9], PATIKA [10], BioPax[11], CellDesigner [12] and Cytoscape [13]. They support different notations for representing these networks other than MIM notation. Another tool, called Pathvisio [14], has added a MIM plug-in[15]. However, Pathvisio does not offer some of the powerful features that are provided by MIMTool which simplify the construction of large MIM diagrams. This is because complete support for MIMs using the Kohn

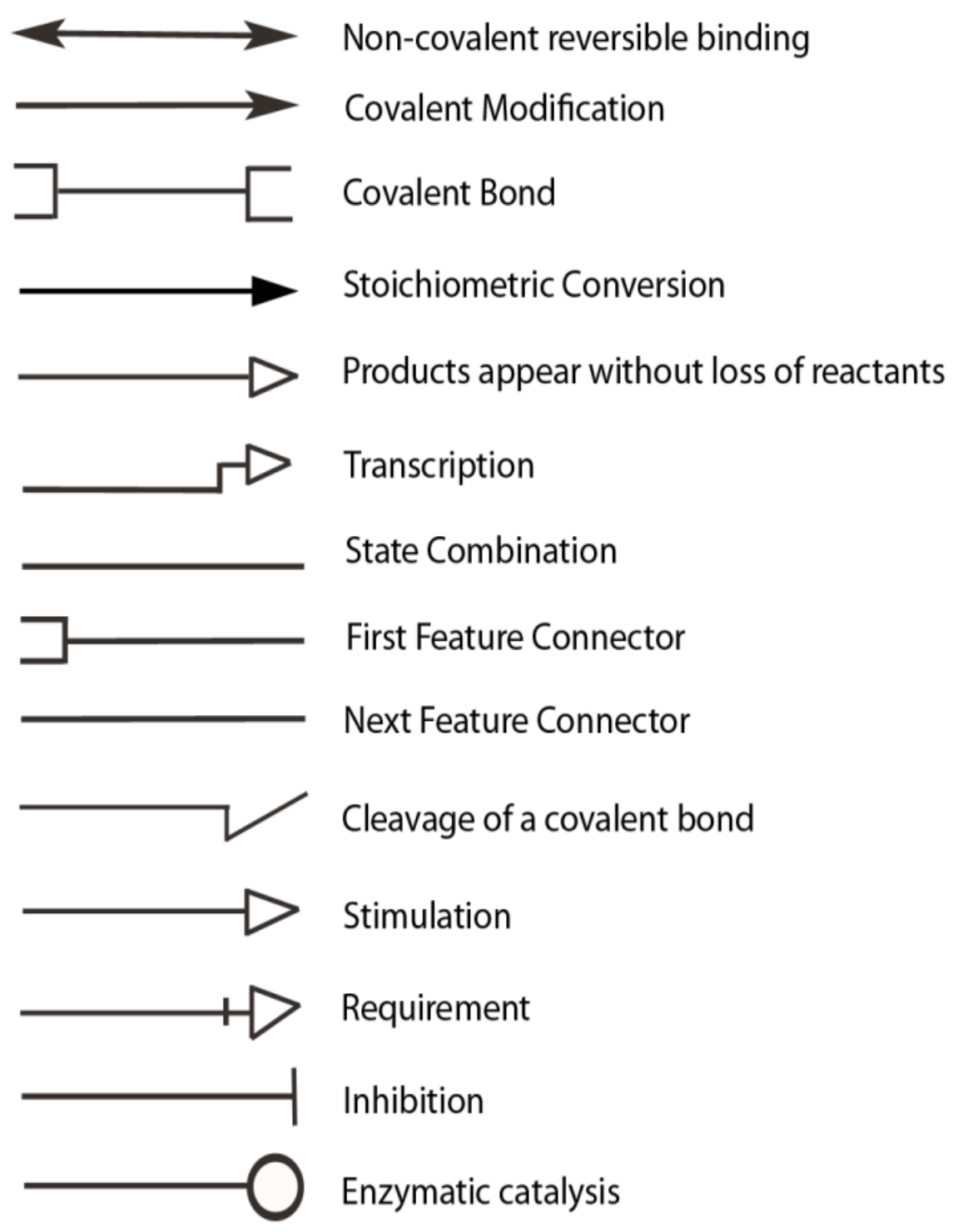

**Figure 1: Interaction Notations of MIM Notation**

notation was not an original design requirement of PathVisio.

There are also 3D models for visualizing protein interaction networks. In recent years, some of the above mentioned tools added 3D visualization plug-ins. For example Cytoscape added 3DScape plug-in that support 3D visualization and layout of Cytoscape networks [16]. There are also techniques that can be utilized in 3D protein visualization system products [17,18]. However since MIM notation is 2D, the focus on this paper is on 2D visualization of protein interaction networks.

For 2D modeling of ordinary graphs, there are tools that offer different layout algorithms such as circular, hierarchical and orthogonal. Interactions in Kohn's notation are drawn orthogonally (i.e. as rectangular paths). Orthogonal drawing algorithms have received a lot of attention because of their application to many real life systems. Some of these include VSLI, database design and circuit design. Most research in this area has focused on drawing of planar orthogonal graphs. Examples include Tamassia [19], Battissa [20] and Fössmeier and Kaufmann [21]. However, the algorithms developed in these works generally apply to simple graphs without a given layout. Even if a layout is given, the algorithm places the nodes (mostly on a grid), and then draws all the interaction lines in batch fashion simultaneously. For our semi-automatic drawing problem, the nodes are already placed on the map. Also, as shown in Figure 1, the nodes and edges in MIM notation have far more complex shapes than the simple node and arc shapes in graphs. Since in our problem, we want to draw a path between two nodes whenever a new interaction is to be created, our drawing problem is closer to edge routing problem among complex shapes. In order to do the routing, we make use of the well-known shortest path problem.

The problem of finding the shortest path is one of the most studied topics of network analysis. One of the reasons is that shortest path problem has many applications in both research and in real life applications. There are some known algorithms to solve the path algorithm like Bellman-Ford's algorithm, Floyd Warshall's algorithm and Dijkstra's algorithm [22]. In our study, we have used Dijkstra's shortest path algorithm. It solves the single source shortest path algorithm in a connected graph where lengths of edges of the graph are non-negative [23]. Since in our drawing problem, we have non-negative weights, we used Dijkstra's algorithm which is faster than the alternative algorithms

## MIM Drawing Problem

In the last decade, a lot of data has been gathered about molecular interactions that regulate cell behavior. These data need to be organized and analyzed. For that reason, there is a need for formalized notations to describe bimolecular interaction networks that are drawn consistently and semantically clear. Kohn's MIM notation and SBGN [24] are examples of such notations; here, we focus specifically on the MIMs. Kohn's notation has the following features: (i) ability to include contingencies, (ii) enabling a user to specify known molecular data like protein modifications and complex formation, (iii) displaying complex set of regulatory network interconnections in rich detail, and (iv) ability to capture different cell types and cell states since it can show the interactions of molecules that exist at the same time and place. These notational features can empower researchers while analyzing protein interaction networks. Also as stated in [25, 26], MIM can facilitate computer simulation, information organization and representation of complex combinatorial schemes [27, 28, 29].

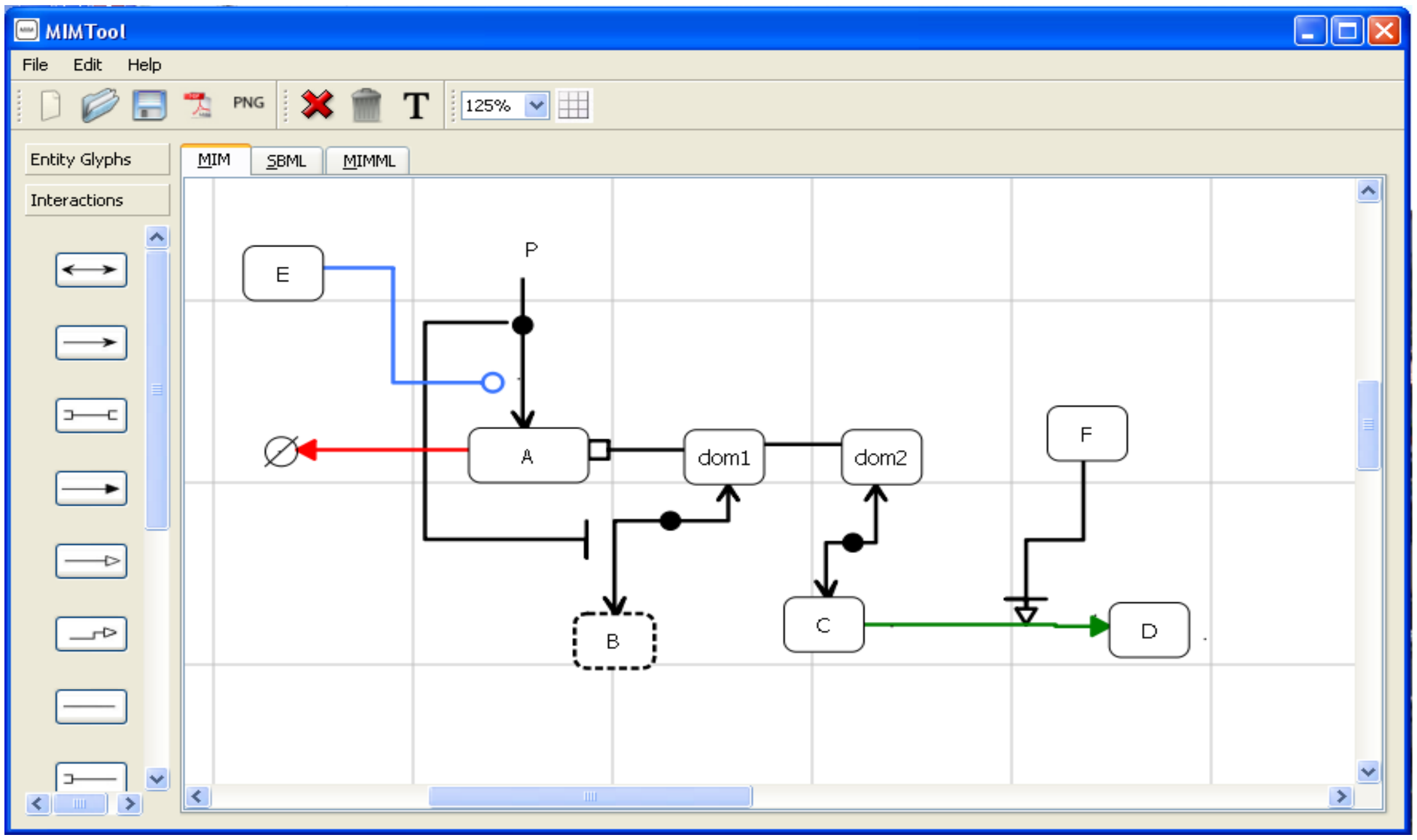

**Figure 2. Snapshot of MIMTool with an example MIM**

Our tool implements the MIM notation as specified in Luna, et al. [7]. In MIM notation, there are two categories of elements. One of them is physical entities which is also known as molecular species. These entities can be elementary or complex. Elementary species are proteins and DNA elements. A protein node is represented as a rectangle with a rounded corner. The second category of elements constitutes molecular interactions between species. These interactions are represented by orthogonal lines with different endings. There are two types of interactions; reactions and contingencies. Reactions operate on molecular species, and they only connect two molecular species together. Contingencies on the other hand are between molecular species and interactions [6]. Figure 1 shows the symbols for interactions.

User friendly, fast and efficient drawings of MIMs are important requirements. Manually drawing a complex molecular interaction map like p53 or p73 can be very time consuming and hard to optimize. Therefore, we have added features to support easy drawing of MIMs both manually and semi-automatically.

## Implementation

### Design Decision for 2D objects

For visualization of 2D MIM notation objects, we have used Qt's QGraphicsView framework. We crated three base classes from this framework; GraphicsScene, GraphicsView and GraphicsItem. GraphicsScene stores and manages graphic items. GraphicsView object visualizes the scene through the main application window. Finally GraphicsItem is the base class of graphical items in the scene.

For MIMTool, we use a scene called a Scene object to visualize the map. For MIM symbols, we use two different graphics items; one for entities and one for interaction lines. These are called DiagramItem and Edge objects respectively.

A main application window called the MainWindow class keeps information about the structure and visual elements of the tool and connects all other classes together.

Scene class object creates a visual scene for MIM and manages all 2D items. GraphicsItems objects are drawn in the scene and GraphicsView object visualizes the scene through the main window.

Besides management of graphical items, Scene object also has some member functions that implement important features of MIMTool. These include saving MIMML and SBML files, saving image files and drawing automatic interaction lines.

As mentioned earlier, MIM notation symbols can be broadly grouped into two categories. The first group constitutes physical entities. These can be considered as nodes of the molecular interaction graphs. DiagramItem class object represents this molecular interaction. The second group constitutes interactions. Edge class object represents molecular interactions in MIM notation. It can be viewed as lines or edges of molecular interactions maps.

As required by the MIM notation, interactions are drawn orthogonally. MIMTool draws the interactions in a rubber-band fashion; i.e. they move together with the diagramItems they are connected to, so that it is easier for a user to manually optimize the positions of the objects. Edge objects are placed on the graph by clicking on the bending points on the scene. This drawing is very flexible. However, it can be hard to do it manually when there are many nodes and interactions existing already on the map. Therefore, we have developed a shortest path based drawing algorithm to automatically do this in an intelligent fashion for the user.

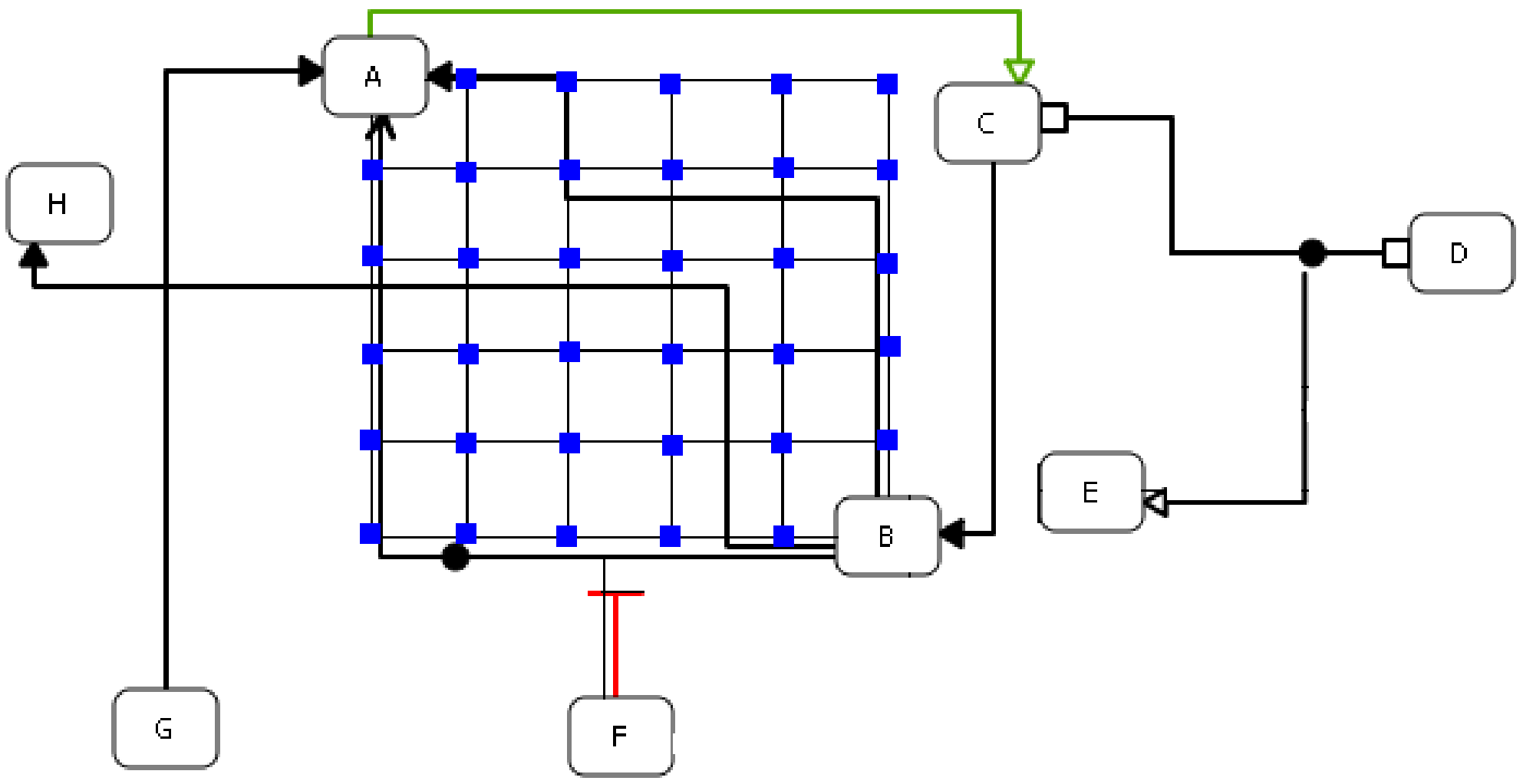

**Figure 3. Screenshot of MIMTool with imaginary grid**

### Features of MIMTool

Features of MIMTool can be summarized as follows:

(i) *Fully interactive*: MIMTool is designed to be flexible for both drawing and successive editing of MIMs. Molecular species can be placed anywhere on the map and dragged anywhere with the mouse. Interaction lines using Manhattan (orthogonal) geometry can be drawn and dragged around with the mouse. When an item is dragged on the scene, all other connected items, whether interactions or molecular species, are also dragged. These highly interactive capabilities give users complete freedom while drawing maps and also make it much easier to introduce changes if necessary.

(ii) *Automatic reaction interaction line drawing*: When a map becomes crowded or large, it can be hard to manually place an interaction line with the mouse. To resolve this difficulty and speed-up the drawing process, an automatic interaction line drawing capability was implemented. This capability enables automatic drawing of an interaction line after two items that to be connected are selected. This orthogonal line placement algorithm was developed by making use of Dijkstra's shortest path algorithm in order to minimize the number of interaction line bends and crossovers.

(iii) *MIMML support:* MIMML is an XML-based storage format developed by Kohn and his group [7]. MIMTool can save layout information of MIMs in MIMML.

(iv) *SBML Export:* MIMTool can export maps as SBML files (Level 2, Version 4). The ability to export SBML files increases the compatibility with many other tools that can import SBML. We used the libSBML [27] library to manage SBML files.

(v) *Code viewing:* There are two tab windows besides the map drawing window. These are read-only text editors for generated MIMML and SBML codes of the drawn maps.

(vi) *Image Export:* MIMTool can save the exported MIMs into image formats for direct visualization. Currently, PDF and PNG formats are supported.

Figure 2 shows an example MIM diagram prepared with MIMTool.

### Shortest Path based Interaction Drawing

In this section, we describe our automatic interaction drawing algorithm in detail. First, we had to adapt MIMs so that they could be used within the context of the Dijkstra's algorithm. We had to decide what will be the nodes and paths used by the algorithm when two nodes are selected for connection by an interaction. To do this, an imaginary grid is placed on the selected nodes. This grid enables us to construct an orthogonal drawing since the paths on the grid are orthogonal. The algorithm has to select one of the paths on the grid.

Figure 3 gives an example with a grid laid out. The size of the grid is chosen as 6x6, meaning a total of 36 nodes were placed on the grid. While choosing the grid size, we take into account the running time of the algorithm and aesthetic appeal. Table-1 lists the running times of different grid sizes on a more complicated example on the same map; finding a path between A and B. The tests were carried out on a Fujtsu-Siemens Amilo M1425 Intel Centrino 1.7GHz, 60GB, 2x256MB notebook. If we increase the number of grids, then the algorithm runs slower. When we set it to 11x11, there are 121 nodes and algorithm slows down. In this case, running time is over 1 second which maybe unappealing to the user. On the other hand, an advantage of this is that when there are more nodes and paths, the algorithm can avoid other items in the scene more easily. However, while avoiding these items, the path start to make bends; hence the total number of bends is increased. If we decrease the grid size, the algorithm runs faster but it cannot avoid most of the items. In a 4x4 grid, it becomes harder to avoid other items in the scene. Therefore we have chosen 6x6 grids. The algorithm can chose the shortest path from 60 line segments.

**Table 1. Running time with different grid sizes**

| Grid size | Running time (secs) |
|---|---|
| 4x4 | 0.18 |
| 5x5 | 0.29 |
| 6x6 | 0.48 |
| 7x7 | 0.73 |
| 9x9 | 0.92 |
| 11x11 | 1.9 |

**Pseudo Code**

Pseudo code of the algorithm is given in Figure 4. Table 2 presents the definitions used in the algorithm. The algorithm starts with a creation of heap data structures since we have used the bidirectional Dijkstra's algorithm with a heap implementation. We start the algorithm both from a start node and an end node simultaneously. Therefore, we create two heaps; one for forward and one for reverse algorithm. We also create a boolean variable called test to alter the main loop when necessary conditions are met.

There are two parameters associated with every node: distance and predecessor. Distance defines the total distance value (or total cost) from the starting node. Predecessor stores the predecessor node for the associated node in the shortest path. In another words, it is the adjacent node with the minimum distance value.

In lines 4 through 6, distance values are initialized. For the forward algorithm, the starting node has the minimum distance with zero. For the reverse algorithm, the end node has the minimum distance with zero. All the other distance values are set to a very large number *M*. Then in lines 7 and 8 predecessor node of the starting node and the end node is set to zero. Starting node and end nodes are inserted into the heaps and the main loop of the algorithm begins in line 11.

**Table 2 : Definitions used in the rectangular drawing algorithm**

| | |
|---|---|
| *G = (N,E)* | graph *G* with node set *N* and edge set *E* |
| *s* | source node |
| *d* | destination node or end node |
| *e(i,j)* | edge from node i to node j |
| *d(i)* | length (cost) of the edge between nodes *i* and *j* |
| *c(i,j)* | distance(total cost) from source node to node *i* |
| *pred(i)* | predecessor of node *i* |
| *cross(i,j)* | checks if a line from node *i* to node *j* would cross any other object in the scene |
| *bend(i,j)* | checks if a line from node i to node j causes bending |
| *adj(i)* | adjacency list of node *i* |
| *test* | Boolean variable that stores ending criteria for the main while loop |
| *CROSSCOST* | cost of crossing (a constant set to 20) |
| *BENDCOST* | cost of bending (a constant set to 20) |

The algorithm starts with the forward algorithm. We select the node with the minimum distance and remove that node from the heap as shown in lines 12 and 13. Then by calling the Update routine, we update the distance values of the adjacent nodes. For each adjacent node, we first update the cost by adding the length of the edge. Then there are two additional costs; cross and bending. Those extra costs are added for aesthetic appeal which will be discussed in next section.

After the distance of the adjacent node is calculated (in line 30 of the Update routine), we check to see if the previous distance of that node is smaller than the currently calculated distance. If this is the case, then we do nothing. However, if the newly calculated distance is smaller, we have two options. If the former distance value is infinity (which basically means that the distance has never been updated for this node), we update the distance value to the new one and insert the node to the heap. If the former distance value is not equal to infinity but is still greater than newly calculated value, then we update the distance value and call the *decreaseKey* function. This function will update the position of the node in the heap according to the new distance value. Finally we set the current node as the predecessor of the adjacent node in line 37. After the forward algorithm is finished for a node, we start the reverse algorithm. We do all

```
Algorithm Rectangular-Draw
  begin
      Create-heap(Hforward) ;
      Create-heap(Hreverse) ;
      test = false ;
      d(i) = M a very large number for each node i in N ;
      d(s) = 0 ;
      d(e) = 0 ;
      pred(s) = 0 ;
      pred(e) = 0 ;
      insert(s,Hforward) ;
      insert(d,Hreverse) ;
      while ( test = false ) do
            node i = find-min(Hforward) ;
            delete-min(Hforward) ;
            for each e(i,j) element of adj(i) do
                  Update(i,j,Hforward) ;
            endfor
            node m = find-min(Hreverse) ;
            delete-min(Hreverse) ;
            for each e(m,j) element of adj(m) do
                  Update(m,j,Hreverse) ;
            endfor
            If ( find-min(Hreverse) is in vertices scanned by Hforward  or
                find-min(Hforward) is in vertices scanned by Hreverse) then
                test = true ;
      endwhile
  end

Routine Update(p,q,heap)
begin
        cost = d(p) + c(p,q);
        if ( cross(p,q) )  then  cost = cost + CROSSCOST ;
        if ( bend(p,q) )  then cost = cost + BENDCOST  ;
        if ( d(q) > cost ) then
              if  ( d(q) = ∞ ) then
                    d(q) = cost ;
                    insert(q,heap) ;
              else
                    d(q) = cost;
                    decreaseKey(cost, p, heap) ;
                    pred(q) = p ;
              endif
        endif
  end
```

**Figure 4: Rectangular drawing algorithm used in MIMTool**

the same calculations with end node as our first minimum distance node. After all adjacent node calculations are finished for reverse algorithm, we do a boolean check. We test if the minimum distance node of forward heap is in *Hreverse* or minimum distance node of reverse heap is in *Hforward*. This checks if a node is reached by both forward and reverse algorithms. If the boolean variable is true, then the loop ends and we start calculating the shortest path.

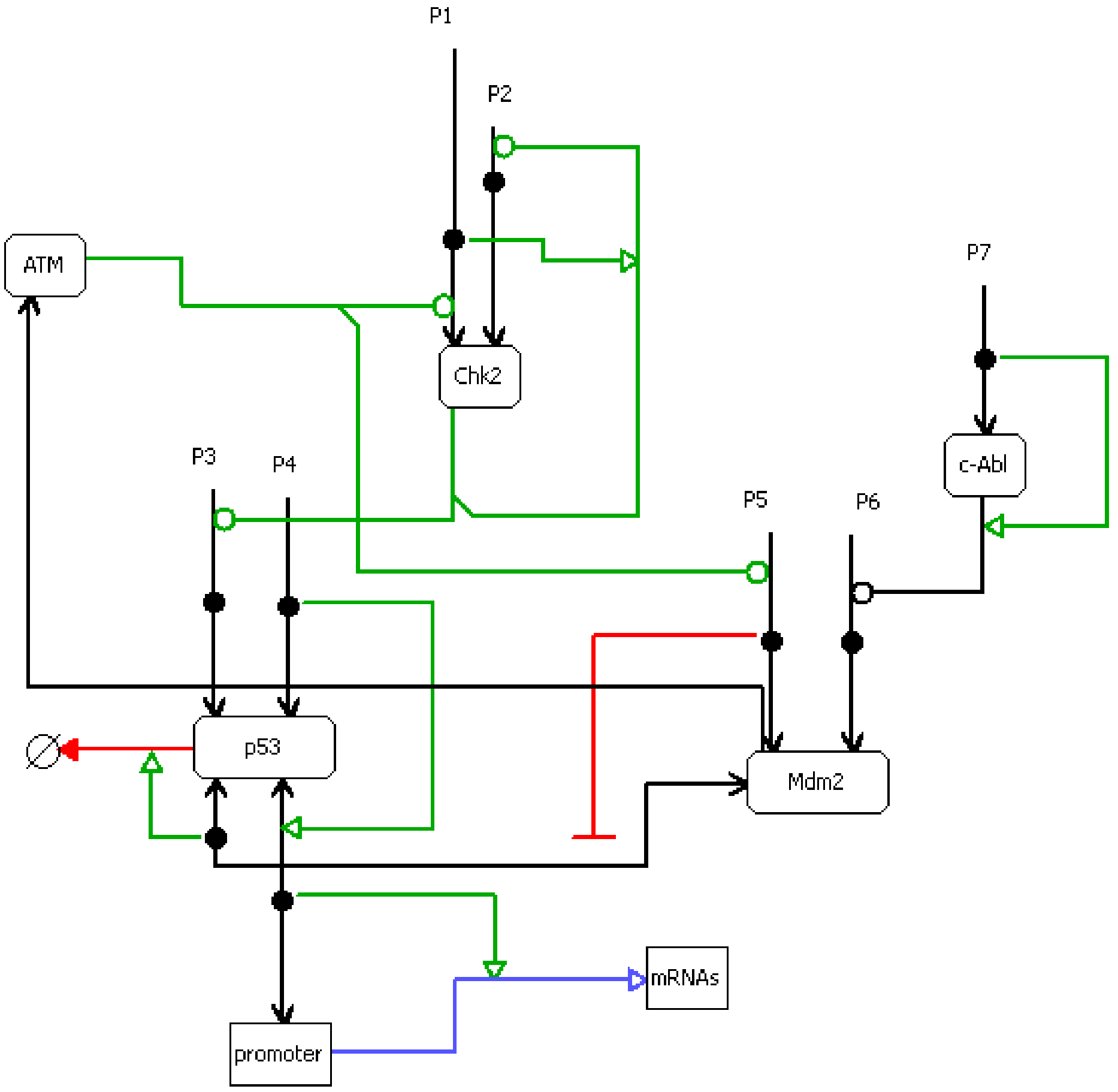

**Figure 5. Example MIM drawn with MIMTool**

We find the shortest path from the node reached by both forward and reverse algorithms. From that node, we get the predecessor nodes from *pred* variable from both forward and reverse algorithms until we reach start and end nodes. Those nodes and edges between those nodes constitute the shortest path.

### Cost function and aesthetic appeal

An issue we had to decide was the length (cost) of a line on the grid. For this decision, we had to take into account our aesthetics constraints; minimization of total number of crossings and minimization of total number of bends. We decided that the length of a line between imaginary nodes should be 5 units if that line does not cross any items. If it crosses an interaction line, the length of the path increases by 20 units. If it crosses a molecular species of any kind, then the length will be increased by 2000 units. Assignment of such a big number can lower the chance of selection of that line as part of the shortest path.

To test if selection of the line creates a bending, we have to check the previous line. In Dijkstra's algorithm, we store information about the previous nodes. Thus, we conduct a test of bending. If the line bends, then length is increased by 20 units, the same as crossing.

## Results and Discussion

### Comparison to existing tools

There are many biomolecular network visualization tools. All of them are important and serve many purposes. However to pinpoint the importance of MIMTool, we will compare tools that are similar to MIMTool. These are: CellDesigner, Cytoscape and Pathvisio. These tools have many important features that MIMTool is yet not capable of doing, for example; connection to online databases, SBML import, etc. However, MIMTool is the first tool to have full support of the complex MIM notation and the MIMML format with features designed to facilitate the drawing of MIMs. These include semi-automatic drawing, code viewing and SBML export. PathVisio has added a MIM plug-in [15]. However, it does not have some key features that is provided by MIMTool, since drawing MIMs is not its primary feature. For example orthogonal drawings of edges are not flexible in PathVisio. Number of bends are limited. However with MIMTool, you can bend edges as many times as you like. Another example is Cytoscape. It supports many different formats. However, it does not support MIM notation. Also it does not support orthogonal edge drawing. It has an orthogonal layout algorithm but it places the nodes in an orthogonal fashion. It does not bend edges. CellDesigner supports orthogonal drawing. CellDesigner can draw diagrams using Kitano's SBGN [24]. SBGN notation is developed by the community of biochemists, modellers and computer scientists to represent biological network models. Like the MIM notation, it is developed to standardize the representation of biological processes and interactions. Even though it is a useful and highly capable notation, SBGN lacks certain key features that MIM possesses. Some of the differences and the relationship between MIM and SBGN have been recently discussed [7].

We note that MIMTool can differentiate itself from existing drawing tools since it was designed and developed especially for the MIM notation For this reason, it provides flexible orthogonal drawing that enables easy MIM drawing and modification.

MIMTool has been tested by our project members and students from Boğaziçi University Computer Engineering (CMPE) and Chemical Engineering (CHE) department. Compared to other tools, MIMTool has been reviewed as simple and easy to use.

## Conclusions

MIMTool can draw Molecular Interaction Maps using Kohn notation. It can be used to draw and edit complicated MIMs. Figure 5 is an example of a MIM drawn by MIMTool. One of our aims for developing MIMTool was to develop a tool that would support Kohn's MIM notation and MIMML and hence facilitate their wider adoption.

MIMTool is a state of-the-art drawing tool for MIMs. It introduces unique features such as object dragging and line drawing with automatic bending and crossover minimization which are not available in most diagram editors for MIMs. MIMTool expands and contributes algorithms and novel drawing techniques to the repertoire of tools that are used to create and edit MIMs.

## Availability and requirements

**Project name:** MIMTool
**Project home page:** http://code.google.com/p/mimtool/
**Operating systems:** Windows, Unix-like (Linux, Mac OSX)
**Programming language:** C++
**Other requirements:** Qt 4.6.3 libraries
**License:** GNU GPL v3.

## Competing interests

The authors declare that they have no competing interests.

## Authors' Information

ME: mine.edes@boun.edu.tr
CO: ozturaca@boun.edu.tr
TH: turkan@prc.boun.edu.tr
AL: augustin@mail.nih.gov
RN: ruthnu@helix.nih.gov

## Authors' contributions

ME designed and implemented the MIMTool and prepared the manuscript. CO coordinated and participated in the design and implementation of the project and edited the manuscript. TH, AL and RN helped in establishing the requirements for MIMTool, participated in the discussions about its design and edited the manuscript.

## Acknowledgements

This work was supported by the Scientific and Technological Research Council of Turkey (TUBITAK) under the grant number 107T382 and in part with Federal funds from the National Cancer Institute, National Institutes of Health, under contract number HHSN261200800001E. The content of this publication does not necessarily reflect the views or policies of the Department of Health and Human Services, nor does mention of trade names, commercial products, or organizations imply endorsement by the U.S. Government. It was also supported in part by the Intramural Research Program of the NIH, National Cancer Institute, Center for Cancer Research.

## Author details

[1] Computational Science and Eng. Program, Bogazici University, Istanbul, Turkey
[2] Dept. of Computer Eng., Bogazici University, Istanbul, Turkey.
[3] Dept. of Chemical Engineering, Polymer Research Ctr., Bogazici University, Istanbul, Turkey.
[4] Lab. of Molecular Pharmacology, Ctr. for Cancer Research, National Cancer Inst., NIH, Bethesda, MD 20892, USA and Dept. of Bioinformatics, Boston University, Boston, MA, USA.
[5] Ctr. for Cancer Research, Nanobiology Program, SAIC, Frederick National Lab for Cancer Research, National Cancer Inst., NIH, Frederick, MD 21702, USA